# Which Football Player Bears Most Resemblance to Messi? A Statistical Analysis

Jiri Mazurek

*Silesian University in Opava*
*Czech Republic*
*mazurek@opf.slu.cz*

**Abstract**: Many pundits and fans ask themselves the same question: Which football player bears most resemblance to Lionel Messi? Is it Chelsea's Eden Hazard? Is it Paulo Dybala, the heir to Messi in the national team of Argentina? Or is the most alike player to Messi someone completely else? In general, the research on the evaluation of players' performances originated in the context of baseball in the USA, but, currently, it is of great importance in almost every team sport on the planet. Specifically, football clubs' managers can use the data on player's similarity when looking for replacement of their players by other, presumably similar ones. Also, the research in the presented direction is certainly interesting both for football pundits and football fans. Therefore, the aim of this study is to answer the question from the title with the use of the statistical analysis based on the data from ongoing league season retrieved from WhoScored (WS) database. WS provides detailed data (up to 24 parameters such as goals scored, the number of assists, shots on goal, passes, dribbles or fouls) for players of TOP 5 European leagues, and ranks them with respect to their overall performance. For this study, 17 parameters (criteria) most relevant for an attacking player were used, and a set of 28 players, candidates to be "most alike to Messi" from WS TOP 100 list were selected. After data normalization and application of a proper metric function the most similar player to Lionel Messi was found.

## Introduction

Many pundits and fans ask themselves the same question: Which current football player bears most resemblance to Messi by his performance on a pitch? Surely, Messi was often compared to Diego Maradona for its physical appearance and dribbling abilities. But who is the most similar player to Lionel Messi now? Is it Chelsea's Eden Hazard? Is it David Silva, who used to play the role of Messi in the Spanish national team? Is it Paulo Dybala, a diminutive Argentinean, who is Messi's natural heir in the national team? Or is the most alike player to Messi someone completely else?

The problem of a statistical comparison of players is not a new one. In the 1970s, George William James pioneered the use of similarity scores for baseball players with the aim to compare current players with players of the past already introduced in the Hall of Fame. Later, similarity scores became a part of *sabermetrics*, the empirical analysis of baseball, see e.g. Lewis (2004), Sullivan (2004) or James (2010). The use of statistics quickly spread into other team sports such as American football, basketball or ice-hockey, see for instance Dean (2007), Bruce (2016) or *blog.war-on-ice.com* for NHL players. In the 2000s, new digital technologies enabled fast acquisition and storage of sports data, and the Internet made it possible to share and retrieve data all over the planet. Now, managing a club without the use of the statistical data about players' performances and health is almost unthinkable. Big clubs employ not only managers, trainers, or physiotherapists, but also statisticians.

In the territory of football, new websites specializing on data acquisition and analysis emerged, such as *WhoScored.com* (WS), *Opta Sports* or *FiveThirtyEight.com*. However, statistical similarity studies of football players are rather rare in literature. One recent exception is the study presented by Caley (2017). The research focused on Neymar and its purpose was to find the most similar player to the PSG star. Out of 9 potential candidates, the most similar to Neymar with respect to three criteria (scoring, progressive passing and share of time spent in the wide forward position) happened to be Real Madrid's Gareth Bale.

As of early 2018, there is no study of this sort focused on Messi. Therefore, the aim of this paper is to answer the question from the title with the use of the statistical analysis based on the detailed players' data in the ongoing 2017-18 league season.

The data were retrieved from WhoScored database. WS provides 24 parameters (criteria) for each player from top 5 leagues (English, French, Spanish, German and Italian) such as number of games (minutes) played, the number of goals scored, the number of assists, passes, long balls and through balls, shots on goals, tackles, fouls, etc. The statistics is divided into *Offensive*, *Defensive* and *Passing* parts. For the analysis in this study, 17 out of 24 parameters were used which were most relevant to a player on a position of a forward or attacking midfielder. Omitted parameters included yellow and red cards (not directly linked to a style of play) and several defensive parameters, such as defensive blocks or own goals, which were zero or close to zero for selected players.

WS ranks all players from the five leagues from 1st place to approximately 1400th place, the ranking is based on players' overall performance. As for the selection of the most likely candidates bearing the most resemblance to Messi, the data of 28 offensive players (and Messi) were analyzed. Forwards and attacking midfielders most alike to Messi were selected from TOP 100 players of WS ranking, with most players operating in Premier League, French Ligue 1 and Spanish La Liga.

The players' data covered their league performances in the 2017/18 season until January 31, which amounted to approximately 20-24 matches for most of the players. Appendix A provides complete data for all selected players and parameters (criteria). For readers' convenience, the best and worst value for each parameter is indicated (in bold) as well.

**Data and method**

Complete data from WS database with explanatory notes are shown in Appendix A. Because different parameters (criteria) in the WS database have different units (and scales), for example pass accuracy is in percent while the number of goals is dimensionless, normalization (feature scaling) of the data was necessary.

Most (13 out of 17) selected criteria can be considered maximization criteria, which means the higher value achieved by a player (such as the number of *Goals scored*), the better performance of a player. Four criteria were minimization criteria (*Offsides*, *Fouls*, *UnschTch*, *Dispossessed*), where the higher values mean the worse performance of a player.

Let $x$ denote a given numerical value for a given player and a criterion (parameter) $i$, and let $f_{MIN}^i$ ($f_{MAX}^i$) denote the minimal (maximal) value over all players for a criterion $i$. The normalization is defined as follows:

Maximization criterion: $\varphi_i(x) = \dfrac{x - f_{MIN}^i}{f_{MAX}^i - f_{MIN}^i}$  (1)

Minimization criterion: $\varphi_i(x) = \dfrac{f_{MAX}^i - x}{f_{MAX}^i - f_{MIN}^i}$  (2)

By formulas (1-2), values $x$ are transformed into values $\varphi(x) \in [0,1]$, so that the best value among all players for a given criterion is 1, and the worst value for a given criterion is set to 0.

After the normalization, distances between players were computed. To measure distance between two points (or functions), a distance or metric functions first introduced by Frechet (1906) are used, but see also more recent works of Linial (2002) or Searcoid (2006).

Let $x = (x_1, ..., x_n) \in R^n$ and $y = (y_1, ..., y_n) \in R^n$ be two points from *n*-dimensional real space. Let $d_p(x, y)$ (also denoted as $L^p(x, y)$ in literature) be the distance between *x* and *y*. Then, the most common distance function is defined as follows:

$$d_p(x, y) = \left( \sum_{k=1}^{n} |x_k - y_k|^p \right)^{1/p} \quad (3)$$

If the parameter *p* = 2 in (3), then the distance function (3) is called *Euclidean metric*. If *p* = 1, the metric is usually called *Manhattan* (or *Taxicab*) *metric*:

$$d_1(x, y) = \sum_{k=1}^{n} |x_k - y_k| \quad (4)$$

Manhattan metric (4) is simple to use and evaluate, therefore it is applied for this study. In our case, each player is considered a point in the 17-dimensional real space, as there are *n* = 17 parameters evaluating each player.

The aim is to evaluate the distance of a given player to Messi, hence relation (4) can be rewritten in the following way:

$$d_1(Messi, y) = \sum_{k=1}^{17} |Messi_k - y_k| \quad (5)$$

In (5), *y* denotes a player from the list of 28 examined players. The player most similar to Messi attains the minimum value of (5).

Because the data gathered for this study also enable to study (linear) relationships between parameters (criteria), figures illustrating some interesting links are provided in the following section along with correlation coefficients and their statistical significance.

**Results**

The distance of all selected players to Lionel Messi is given in Table 1. The player closest to Messi is Phillipe Coutinho (since January 2018 *FC Barcelona*, previously *FC Liverpool*), followed by Eden Hazard (*Chelsea*), Florian Thauvin (*Ol. Marseille*) and Paulo Dybala (*Juventus*). The result of Hazard and Dybala corroborates opinion of many football pundits, while Thauvin's 3rd place is rather unexpected. Also, the results confirm that David Silva, the player once considered "Spanish Messi", has a different style of play.

In the case of Coutinho, for the analysis his performances in FC Liverpool dress only were used, as this study employed the data up to January 31, and Coutinho played only a few minutes for Barcelona up to this date. In particular, Coutinho was close to Messi in assists per game (M: 0.43, C: 0.43), pass accuracy (80% vs 79%), aerials won (0.1 vs 0.1), key passes (2.7 vs 2.9) or through balls (0.5 vs 0.4). Perhaps the largest difference between the two was found in goals per game (0.95 vs 0.5).

Moreover, the data from Appendix A revealed some overall interesting relationships and patterns, which are illustrated via Figure 1 to Figure 9. In Figures 1-9, every player is represented by one data point with his name as a label.

Figure 1 shows the link between goals scored per game and assists per game. The closer is a player to the upper-right corner, the better performance. In this aspect, Neymar is clearly the best.

Figure 2 illustrates the link between shots per game and goals per game. Players are more or less aligned along the diagonal running from the bottom-left corner to the upper-right corner, as the more shots usually mean more goals, and this relationship is close to linear. One player, who sticks out, is Cristiano Ronaldo, who was not able to convert shots into goals as often as his colleagues. On contrary, players above the diagonal have above average conversion rate, see Neymar and Immobile.

Figure 3 reflects passing abilities of players in terms of the number of average passes per game versus pass accuracy in %. In these aspects of play, David Silva is the best.

Figure 4 takes a closer look at lost balls (*dispossessed*) and bad control of a ball (*UnschTch*). Somewhat surprisingly, Alexis Sanchez and Neymar dominate the categories.

Figure 5 provides comparison of average number of passes and key passes per game. Again, players are aligned along the diagonal. Obviously, the more passes means also the more key passes. Neymar and De Bruyne lead the categories, while David Silva is a player with a lot of passes, but relatively low number of these were key passes.

Figure 6 shows average number of dribbles delivered and fouls received per game. Unsurprisingly, Messi, Hazard and Fekir are good, but Neymar is the true King of Dribbles.

Figure 7 takes a look at defensive aspects of play; average number of tackles per game is compared with the number of fouls per game. Clearly, Nabil Fekir is a player who takes his defensive duties very seriously. Meanwhile, Eden Hazard does not seem to care much about defending...

Figure 8 focuses on aerials won and offsides per game. While Cristiano Ronaldo excels in both activities, Romelu Lukaku is the King of the Air and Luis Suarez the King of Offsides (likely to Ernesto Valverde's disenchantment).

The last Figure 9 deals with the passing abilities of players again, in terms of average number of crosses and long balls per game. Hazard, Coutinho or Luis Alberto are true masters of crossing, but the best, and almost off-chart, is Kevin De Bruyne.

Furthermore, the correlation matrix of all variables revealed interesting (linear) relationships. Among the most correlated pairs of parameters were: *average passes-key passes* (Pearson's $\rho = 0.80^{***}$), *dispossessed-dribbling* ($0.78^{***}$), *dribbling-fouled* ($0.77^{***}$), or *through balls-key passes* ($0.73^{***}$), where "***" means statistically significant at $p = 0.01$ level.

The first relationship is shown in Figure 10, the blue line is the corresponding linear trend. Players above the blue line are dispossessed of the ball more often than an average player, and vice versa. In other words, it is harder to dispossess Hazard or Neymar, than Messi or Sanchez.

**Table 1.** Players' distances to Messi.

| Rank | Player | Dist. | Rank | Player | Dist. | Rank | Player | Dist. |
|---|---|---|---|---|---|---|---|---|
| 1 | Coutinho | 3.769 | 11 | A. Sanchez | 5.050 | 21 | C. Ronaldo | 6.000 |
| 2 | Hazard | 4.069 | 12 | L. Alberto | 5.126 | 22 | N. Fekir | 6.067 |
| 3 | Thauvin | 4.140 | 13 | Immobile | 5.145 | 23 | De Bruyne | 6.414 |
| 4 | Dybala | 4.254 | 14 | Mertens | 5.147 | 24 | D. Silva | 6.526 |
| 5 | Aspas | 4.621 | 15 | Di Maria | 5.247 | 25 | Aubameyang | 6.548 |
| 6 | Aguero | 4.705 | 16 | Kane | 5.363 | 26 | Firmino | 6.826 |
| 7 | Salah | 4.849 | 17 | Mahrez | 5.535 | 27 | Mariano | 7.246 |
| 8 | Sterling | 4.858 | 18 | L. Suarez | 5.725 | 28 | Lukaku | 7.489 |
| 9 | Mbappe | 4.910 | 19 | Cavani | 5.744 | | | |
| 10 | Neymar | 4.945 | 20 | Griezmann | 5.788 | | | |

Source: author.

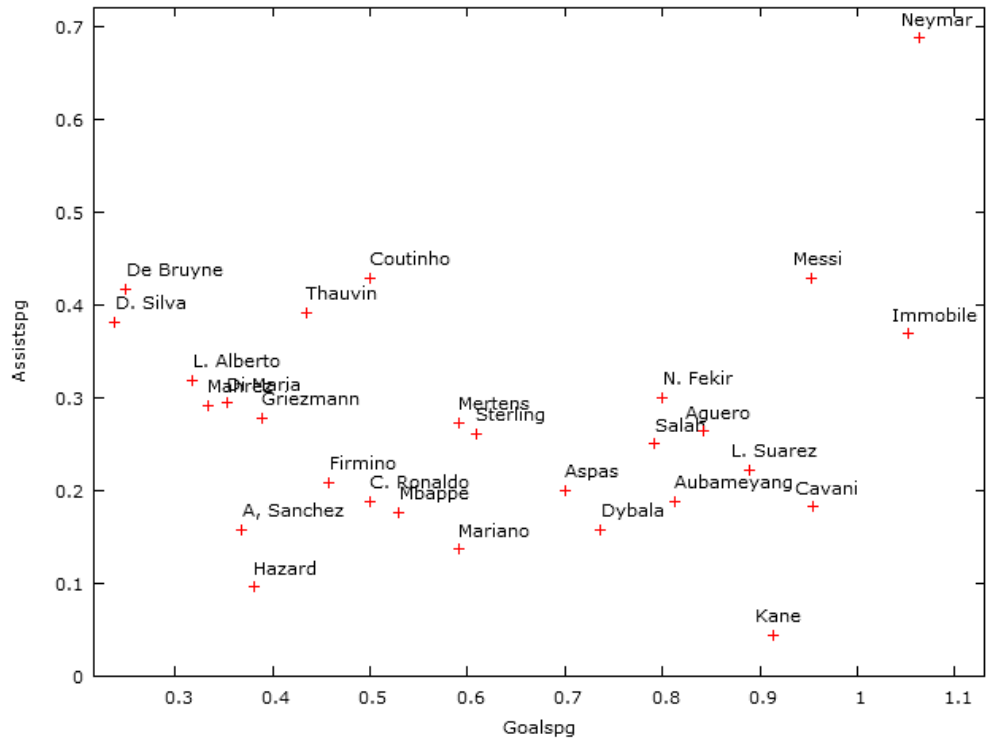

**Figure 1.** Goals scored per game versus assists per game. Source: author.

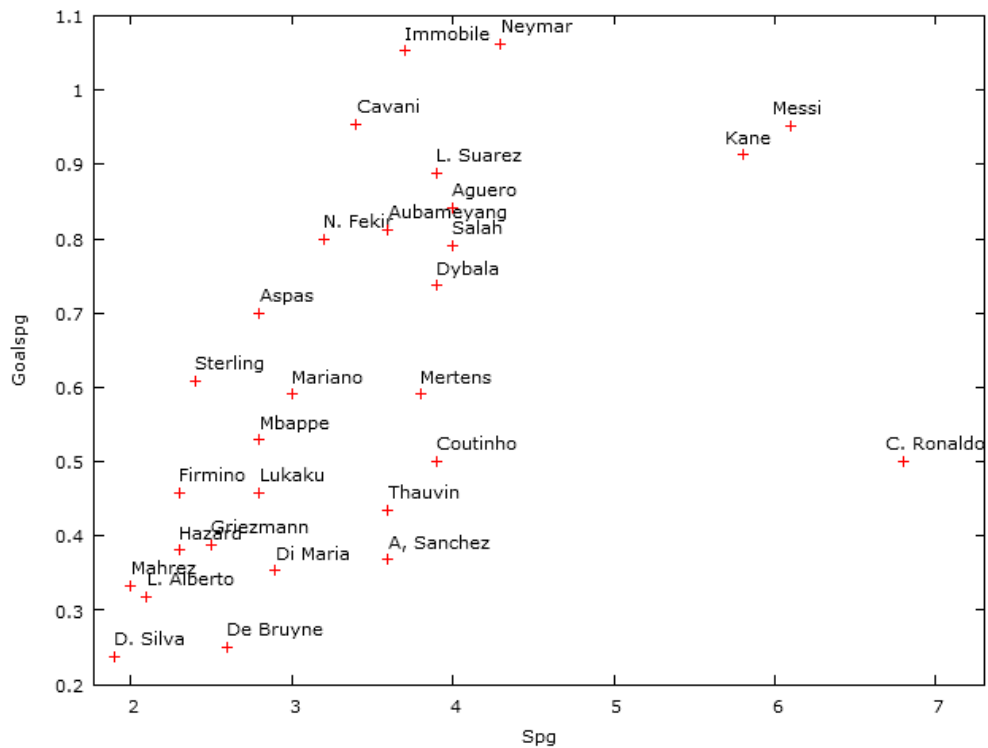

**Figure 2.** Shots per game versus goals per game. Source: author.

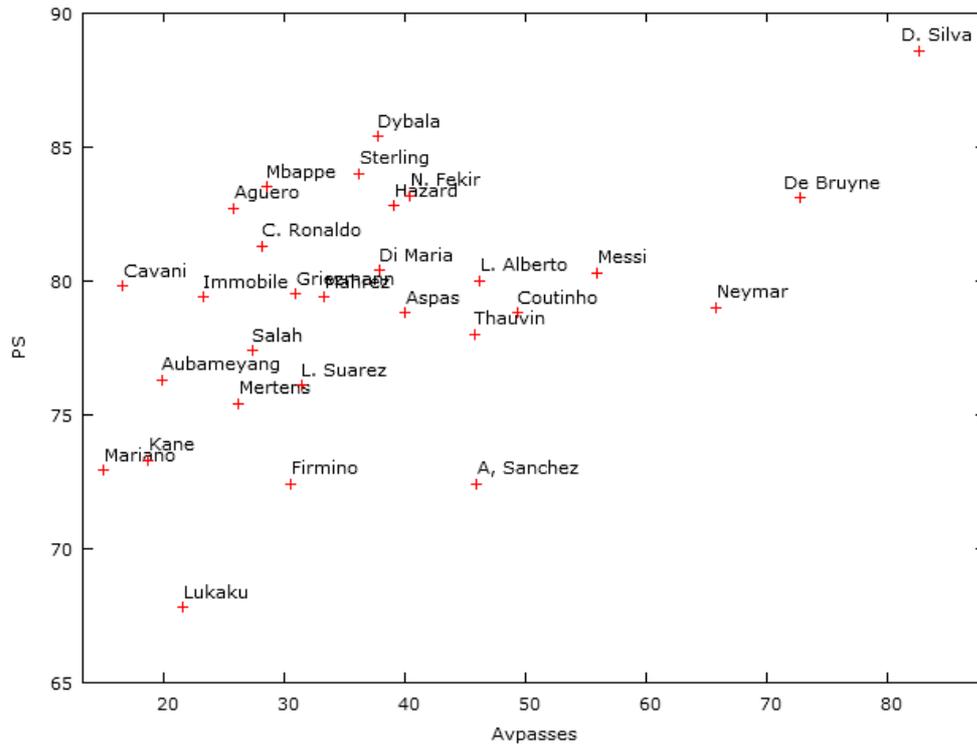

**Figure 3.** The number of average passes per game versus pass accuracy in %. Source: author.

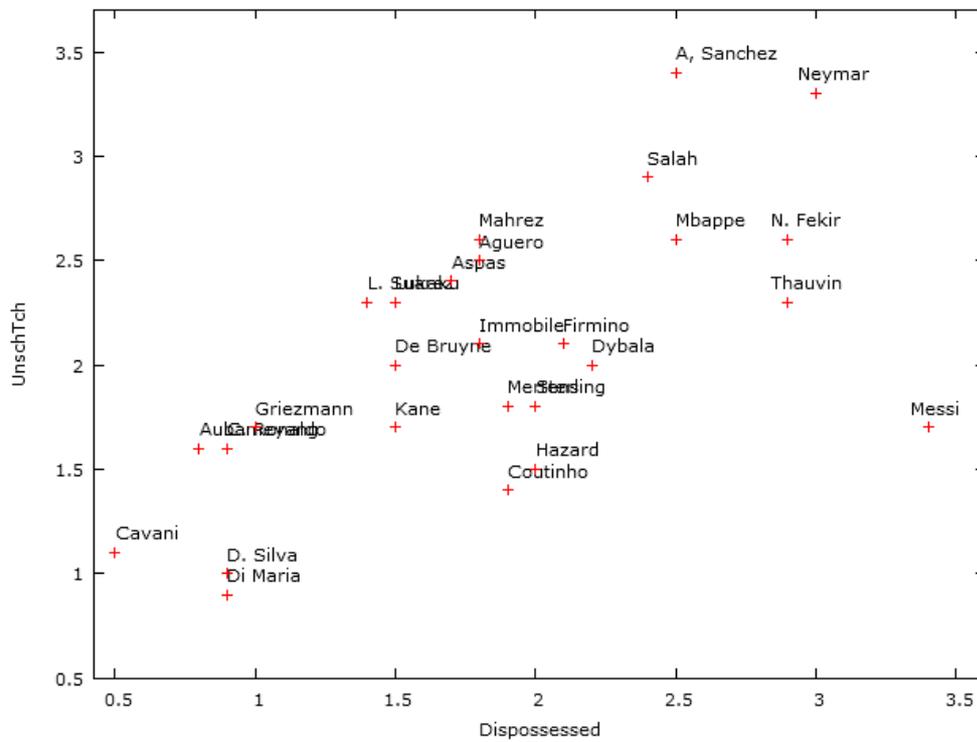

**Figure 4.** Average number of lost balls (dispossessed) per game versus bad control of a ball per game. Source: author.

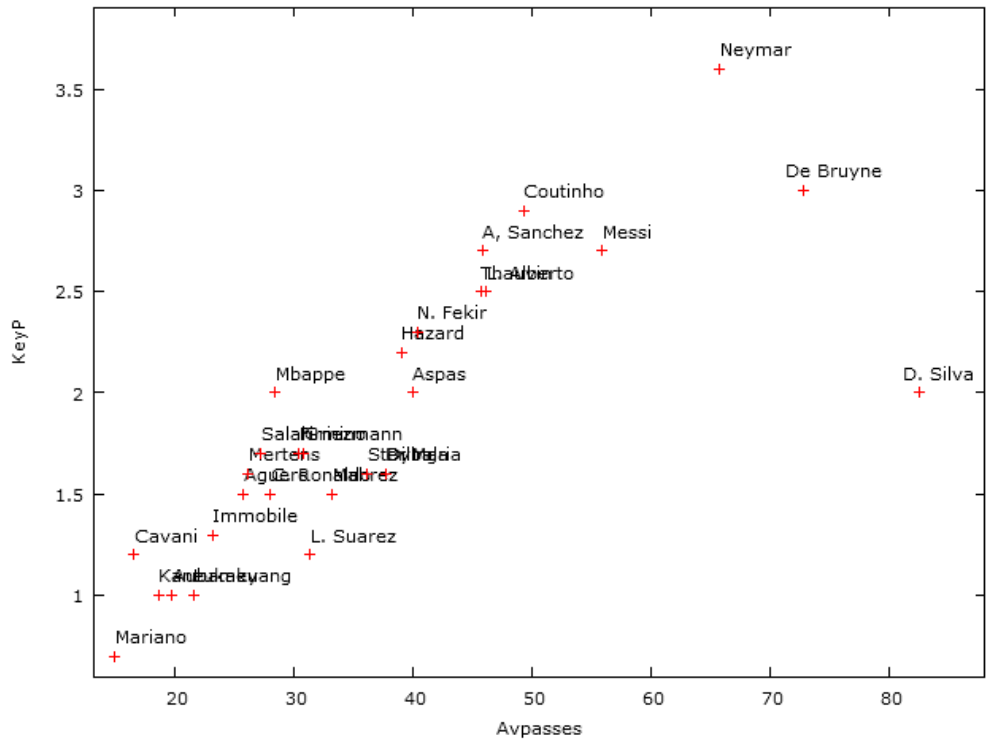

**Figure 5.** Average number of passes per game versus key passes per game. Source: author.

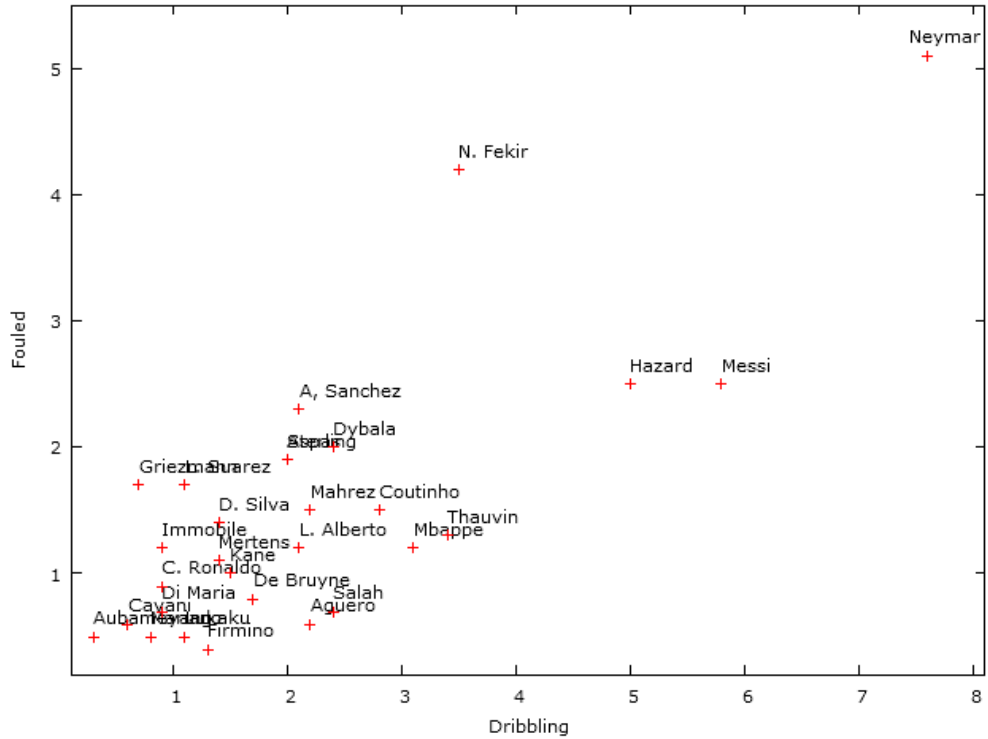

**Figure 6.** Average number of dribbles per game versus the number of fouls received per game. Source: author.

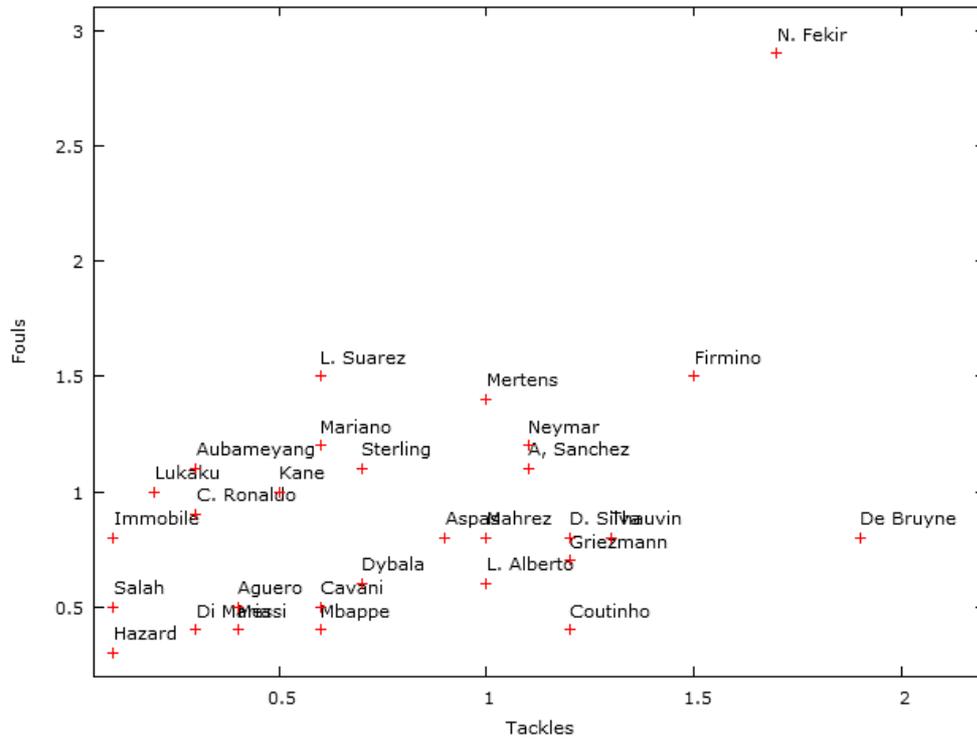

**Figure 7.** Average number of tackles per game versus the number of fouls per game. Source: author.

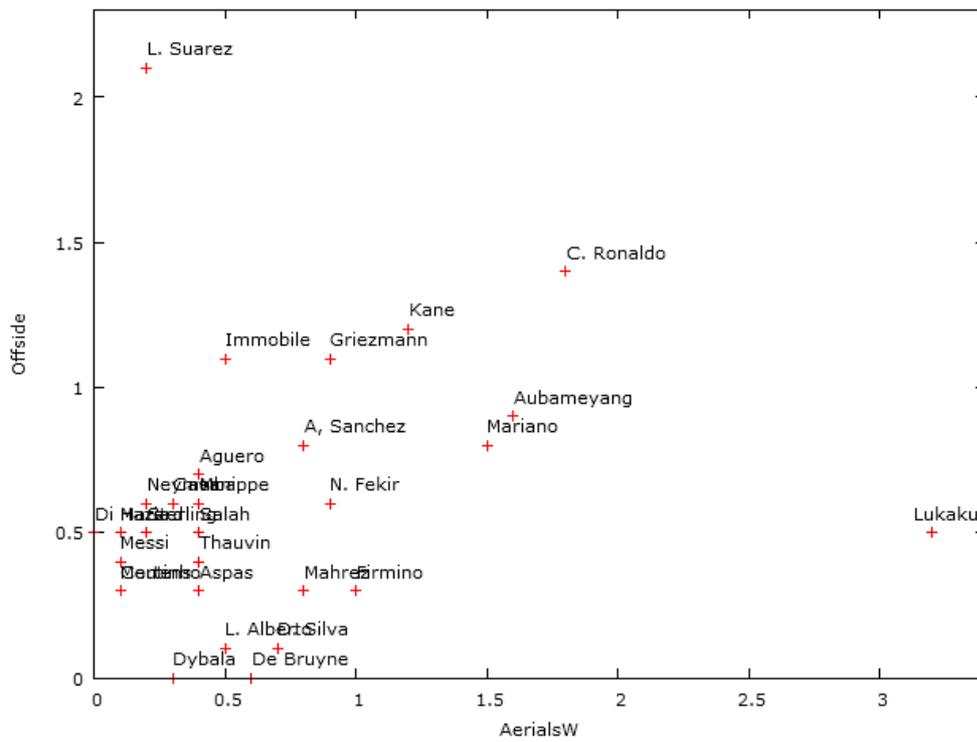

**Figure 8.** Average number of aerials won per game versus the number of offsides per game. Source: author.

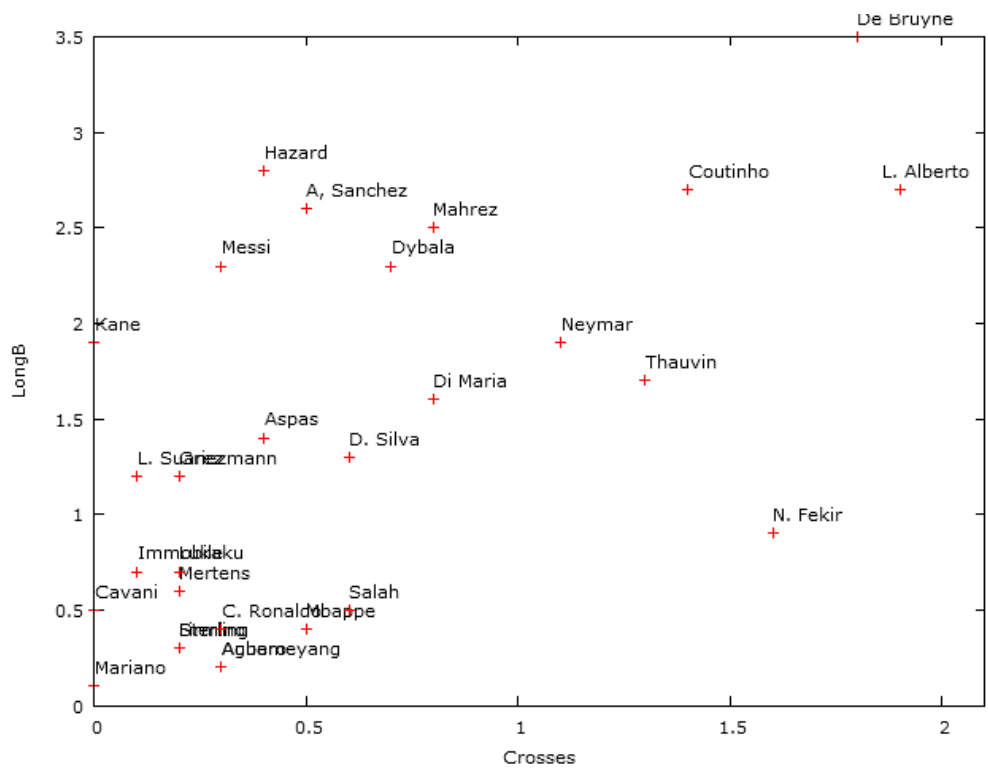

**Figure 9.** Average number of crosses per game versus average number of long balls per game. Source: author.

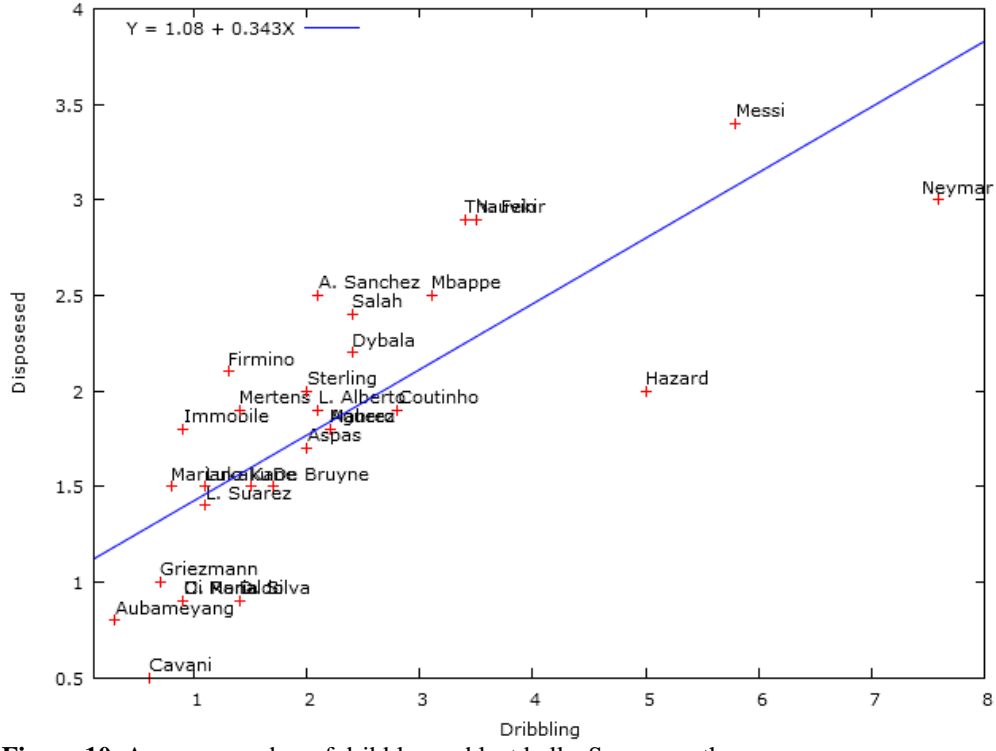

**Figure 10.** Average number of dribbles and lost balls. Source: author.

## Conclusions

In this paper, performances of 29 top players, forwards and attacking midfielders, from TOP 5 European leagues were analyzed with the aim to find a player statistically most similar to Messi. The result is somewhat surprising and "juicy": this player is Phillipe Coutinho, the player, who just happened to be signed in the 2018 winter transfer window by... Futbol Club Barcelona, the home side of Lionel Messi himself! Hence, fans of FC Barcelona may now enjoy not only Lionel Messi, but also his "twin", Coutinho.

As to address a question likely posed by Cristiano Ronaldo's fans, who is the most similar to CR7 (from the set of players in this study), here is the answer: P.-A. Aubameyang (distance 2.29), followed by Harry Kane (3.01) and Antoine Griezmann (3.08).

Neymar happens to be the most distant player to Cristiano Ronaldo. But more importantly, Neymar is currently the best player in the World according to WS rankings, and the data used in this study seem to corroborate this claim. As shown in graphs in the previous section, his present goal scoring rate, assist providing and dribbling skills are unmatched in the football world.

Further analysis of similarity among players may focus on other players, or it can include larger groups of players. The research on this topic could be of great importance for football club managers who look for replacement of their players by other (similar) players. Also, the research in the presented direction is certainly interesting for both football pundits and football fans.

## Acknowledgments

This paper was supported by the Ministry of Education, Youth and Sports Czech Republic within the Institutional Support for Long-term Development of a Research Organization in 2018.

# Appendix A

| Player | Games | Goals | Assists | Spg | PS% | AerW | Dribbling | Fouled | Offside | Disp | Unsch Tch |
|---|---|---|---|---|---|---|---|---|---|---|---|
| Messi | 21 | 20 | 9 | 6.1 | 80.3 | 0.1 | 5.8 | 2.5 | 0.4 | **3.4** | 1.7 |
| Neymar | 16 | 17 | **11** | 4.3 | 79 | 0.2 | **7.6** | **5.1** | 0.6 | 3 | 3.3 |
| N. Fekir | 20 | 16 | 6 | 3.2 | 83.2 | 0.9 | 3.5 | 4.2 | 0.6 | 2.9 | 2.6 |
| Coutinho | **14** | 7 | 6 | 3.9 | 78.8 | 0.1 | 2.8 | 1.5 | 0.3 | 1.9 | 1.4 |
| De Bruyne | **24** | 6 | 10 | 2.6 | 83.1 | 0.6 | 1.7 | 0.8 | **0** | 1.5 | 2 |
| L. Suarez | 18 | 16 | 4 | 3.9 | 76.1 | 0.2 | 1.1 | 1.7 | **2.1** | 1.4 | 2.3 |
| Thauvin | 23 | 10 | 9 | 3.6 | 78 | 0.4 | 3.4 | 1.3 | 0.4 | 2.9 | 2.3 |
| Mertens | 22 | 13 | 6 | 3.8 | 75.4 | 0.1 | 1.4 | 1.1 | 0.3 | 1.9 | 1.8 |
| Kane | 23 | **21** | 1 | 5.8 | 73.3 | 1.2 | 1.5 | 1 | 1.2 | 1.5 | 1.7 |
| Cavani | 22 | **21** | 4 | 3.4 | 79.8 | 0.3 | **0.6** | 0.6 | 0.6 | **0.5** | 1.1 |
| Aguero | 19 | 16 | 5 | 4 | 82.7 | 0.4 | 2.2 | 0.6 | 0.7 | 1.8 | 2.5 |
| Dybala | 19 | 14 | 3 | 3.9 | 85.4 | 0.3 | 2.4 | 2 | **0** | 2.2 | 2 |
| Mbappe | 17 | 9 | 3 | 2.8 | 83.5 | 0.4 | 3.1 | 1.2 | 0.6 | 2.5 | 2.6 |
| Hazard | 21 | 8 | 2 | 2.3 | 82.8 | 0.1 | 5 | 2.5 | 0.5 | 2 | 1.5 |
| Sterling | 23 | 14 | 6 | 2.4 | 84 | 0.2 | 2 | 1.9 | 0.5 | 2 | 1.8 |
| D. Silva | 21 | **5** | 8 | **1.9** | **88.6** | 0.7 | 1.4 | 1.4 | 0.1 | 0.9 | 1 |
| Salah | 24 | 19 | 6 | 4 | 77.4 | 0.4 | 2.4 | 0.7 | 0.5 | 2.4 | 2.9 |
| Mahrez | 24 | 8 | 7 | 2 | 79.4 | 0.8 | 2.2 | 1.5 | 0.3 | 1.8 | 2.6 |
| L. Alberto | 22 | 7 | 7 | 2.1 | 80 | 0.5 | 2.1 | 1.2 | 0.1 | 1.9 | 1.4 |
| Immobile | 19 | 20 | 7 | 3.7 | 79.4 | 0.5 | 0.9 | 1.2 | 1.1 | 1.8 | 2.1 |
| Firmino | 24 | 11 | 5 | 2.3 | 72.4 | 1 | 1.3 | **0.4** | 0.3 | 2.1 | 2.1 |
| Lukaku | 24 | 11 | 5 | 2.8 | **67.8** | **3.2** | 1.1 | 0.5 | 0.5 | 1.5 | 2.3 |
| Aspas | 20 | 14 | 4 | 2.8 | 78.8 | 0.4 | 2 | 1.9 | 0.3 | 1.7 | 2.4 |
| C. Ronaldo | 16 | 8 | 3 | **6.8** | 81.3 | 1.8 | 0.9 | 0.9 | 1.4 | 0.9 | 1.6 |
| Mariano | 22 | 13 | 3 | 3 | 72.9 | 1.5 | 0.8 | 0.5 | 0.8 | 1.5 | 2 |
| A. Sanchez | 19 | 7 | 3 | 3.6 | 72.4 | 0.8 | 2.1 | 2.3 | 0.8 | 2.5 | **3.4** |
| Griezmann | 18 | 7 | 5 | 2.5 | 79.5 | 0.9 | 0.7 | 1.7 | 1.1 | 1 | 1.7 |
| Di Maria | 17 | 6 | 5 | 2.9 | 80.4 | **0** | 0.9 | 0.7 | 0.5 | 0.9 | **0.9** |
| Aubameyang | 16 | 13 | 3 | 3.6 | 76.3 | 1.6 | 0.3 | 0.5 | 0.9 | 0.8 | 1.6 |

| Player | KeyP | AvPasses | Crosses | LongB | ThruB | Tackles | Fouls | Goals pg | As pg |
|---|---|---|---|---|---|---|---|---|---|
| Messi | 2.7 | 55.9 | 0.3 | 2.3 | 0.5 | 0.4 | 0.4 | 0.95 | 0.43 |
| Neymar | **3.6** | **65.8** | 1.1 | 1.9 | **1.2** | 1.1 | 1.2 | **1.06** | **0.69** |
| N. Fekir | 2.3 | 40.4 | 1.6 | 0.9 | 0.2 | 1.7 | **2.9** | 0.80 | 0.30 |
| Coutinho | 2.9 | 49.3 | 1.4 | 2.7 | 0.4 | 1.2 | 0.4 | 0.50 | 0.43 |
| De Bruyne | 3 | 72.8 | **1.8** | 3.5 | 0.3 | **1.9** | 0.8 | 0.25 | 0.42 |
| L. Suarez | 1.2 | 31.4 | 0.1 | 1.2 | 0.2 | 0.6 | 1.5 | 0.89 | 0.22 |
| Thauvin | 2.5 | 45.7 | 1.3 | 1.7 | 0.3 | 1.3 | 0.8 | 0.43 | 0.39 |
| Mertens | 1.6 | 26.2 | 0.2 | 0.6 | 0.3 | 1 | 1.4 | 0.59 | 0.27 |
| Kane | 1 | 18.7 | **0** | 1.9 | 0.1 | 0.5 | 1 | 0.91 | **0.04** |
| Cavani | 1.2 | 16.6 | **0** | 0.5 | 0.1 | 0.6 | 0.5 | 0.95 | 0.18 |
| Aguero | 1.5 | 25.8 | 0.3 | 0.2 | 0.2 | 0.4 | 0.5 | 0.84 | 0.26 |
| Dybala | 1.6 | 37.7 | 0.7 | 2.3 | **0** | 0.7 | 0.6 | 0.74 | 0.16 |
| Mbappe | 2 | 28.5 | 0.5 | 0.4 | 0.2 | 0.6 | 0.4 | 0.53 | 0.18 |
| Hazard | 2.2 | 39.1 | 0.4 | 2.8 | **0** | **0.1** | **0.3** | 0.38 | 0.10 |
| Sterling | 1.6 | 36.2 | 0.2 | 0.3 | 0.2 | 0.7 | 1.1 | 0.61 | 0.26 |
| D. Silva | 2 | 82.6 | 0.6 | 1.3 | 0.1 | 1.2 | 0.8 | **0.24** | 0.38 |
| Salah | 1.7 | 27.3 | 0.6 | 0.5 | 0.1 | **0.1** | 0.5 | 0.79 | 0.25 |
| Mahrez | 1.5 | 33.3 | 0.8 | 2.5 | 0.2 | 1 | 0.8 | 0.33 | 0.29 |
| L. Alberto | 2.5 | 46.2 | 1.9 | 2.7 | 0.4 | 1 | 0.6 | 0.32 | 0.32 |
| Immobile | 1.3 | 23.2 | 0.1 | 0.7 | **0** | 0.1 | 0.8 | 1.05 | 0.37 |
| Firmino | 1.7 | 30.5 | 0.2 | 0.3 | 0.1 | 1.5 | 1.5 | 0.46 | 0.21 |
| Lukaku | 1 | 21.6 | 0.2 | 0.7 | 0.1 | 0.2 | 1 | 0.46 | 0.21 |
| Aspas | 2 | 40 | 0.4 | 1.4 | 0.1 | 0.9 | 0.8 | 0.70 | 0.20 |
| C. Ronaldo | 1.5 | 28.1 | 0.3 | 0.4 | 0.1 | 0.3 | 0.9 | 0.50 | 0.19 |
| Mariano | **0.7** | **15** | **0** | **0.1** | 0.1 | 0.6 | 1.2 | 0.59 | 0.14 |
| A. Sanchez | 2.7 | 45.9 | 0.5 | 2.6 | 0.4 | 1.1 | 1.1 | 0.37 | 0.16 |
| Griezmann | 1.7 | 30.9 | 0.2 | 1.2 | 0.3 | 1.2 | 0.7 | 0.39 | 0.28 |
| Di Maria | 1.6 | 37.8 | 0.8 | 1.6 | 0.3 | 0.3 | 0.4 | 0.35 | 0.29 |
| Aubameyang | 1 | 19.8 | 0.3 | 0.2 | **0** | 0.3 | 1.1 | 0.81 | 0.19 |

**Source**: WhoScored (2018).

**Notes**:
Games... Games played.
Goals... Goals scored.
Assists... Assists provided.
SpG... Shots per game.
PS%... Pass accuracy.
AerW... Aerials won per game.
Dribbling... Number of dribbles per game.
Fouled... Number of times fouled per game.
Offside... Offsides per game.
Disp... Number of times dispossessed per game.
UnchTch... Number of times of bad control of a ball per game.
KeyP... Key passes per game.
AvPass... Number of passes per game.
Crosses... Number of crosses per game.
LonB... Number of long balls per game.
ThruB... Number of through balls per game.
Tackles... Number of tackles balls per game.
Fouls... Number of fouls balls per game.
Goals pg... Goals per game.
As pg... Assists per game.